\documentclass[conference]{IEEEtran}
\ifCLASSINFOpdf
   \usepackage[pdftex]{graphicx}
   \DeclareGraphicsExtensions{.pdf,.jpeg,.png}
\else
\fi
\begin{document}
%
\title{A Compact Model for SiC Power MOSFETs\\ for Large Current and High Voltage\\ Operation Conditions}

\author{\IEEEauthorblockN{Cristino Salcines}
\IEEEauthorblockA{University of Stuttgart\\
Stuttgart, Germany\\
Email: salcines@ilh.uni-stuttgart.de
}
\and
\IEEEauthorblockN{Sourabh Khandelwal}
\IEEEauthorblockA{Macquarie University\\
Sydney, Australia\\
Email: sourabh.khandelwal@mq.edu.au}
\and
\IEEEauthorblockN{Ingmar Kallfass}
\IEEEauthorblockA{University of Stuttgart\\
Stuttgart, Germany\\
}}


%


\maketitle

\begin{abstract}
This work presents a physics based compact model for SiC power MOSFETs that accurately describes the I-V characteristics up to large voltages and currents. Charge-based formulations accounting for the different physics of SiC power MOSFETs are presented. The formulations account for the effect of the large SiC/SiO2 interface traps density characteristic of SiC MOSFETs and its dependence with temperature. The modeling of interface charge density is found to be necessary to describe the electrostatics of SiC power MOSFETs when operating at simultaneous high current and high voltage regions. The proposed compact  model accurately fits the measurement data extracted of a 160 milli ohms, 1200V SiC power MOSFET in the complete IV plane from drain-voltage $V_d$ = 5mV up to 800 V and current ranges from few mA to 30 A.
\end{abstract}


%
\IEEEpeerreviewmaketitle

\section{Introduction}
Silicon Carbide (SiC)  is gradually penetrating the power electronics market \cite{tiwari2016sic, chen20131200, imaizumi2014characteristics}. The fast switching speeds, low on-resistance and superior temperature handling of SiC power MOSFETs presents this technology as a good candidate to replace Silicon in the medium voltage range of hundreds to thousands of volts. The switching speeds and threshold voltage instability of SiC power MOSFETs bring however new challenges for circuit designers and hinders the process of unlocking the full potential of this promising technology \cite{wada2018switching, liu2020characterization}. 
To exploit full potential of SiC technology accurate circuit simulations are a necessity. Simulations save iterative design costs and speed up the overall design process. To achieve accurate circuit simulations, it is essential to provide circuit designers with accurate and convergence-robust transistor models. The development of transistor models that can reproduce the static and dynamic behavior of SiC power MOSFETs play a crucial role in the path to achieve reliable and accurate circuit simulations.

SiC modeling research has progressed from previous empirical models \cite{mukunoki2018improved, mudholkar2013datasheet} to physics-based surface-potential-based models \cite{albrecht2020iterative, nakamura2016simulation}. Silicon Carbide power MOSFETs suffer from a reduced charge carrier mobility in the inversion charge functioning as channel due to a high density of charge traps in the SiC/SiO2 interface. The temperature dependence of the traps density in the SiC/SiO2 interface incurs a reduction of the threshold voltage of the device at high temperatures. Such phenomena combined with the self-heating of the transistor affect the $\frac{dV}{dt}$ and the $\frac{dI}{dt}$ of the power switch, key parameters that define the switching losses of power transistors. Some efforts have been put in improving the accuracy of SiC transistor models by accounting for these effects. However, so far no models have been reported on the modeling of SiC power MOSFETs in the high voltage high current operation region. In this work, we propose a physics-based model for SiC MOSFETs which is valid in all regions of device operations. We have developed new formulations based on the silicon MOSFET compact models \cite{bucher1996accurate, chauhan2013bsim6} which account for the SiC MOSFET specific physics. With the developed model we report for the first time, how the behavior of interface traps w.r.t local temperature is critical to model the high current high voltage regions of I-V plane.

\section{Characterization of SiC MOSFET}
We have selected a 160 milli Ohms/1200V commercially available SiC power MOSFET as device under test (DUT) to develop and the transistor model. Our goal is to develop a model valid for the full I-V plane. One of the challenges is that data-sheet typically do not include characterization plots covering both low and high current regions. Typically a very limited I-V graphs are given. For example to fit the trans-conductance of the power MOSFET in the extreme high voltage region, no information is given in the datasheet. Measurement equipment manufacturers also currently do not have commercial instruments capable of measurnig the high current region. Therefore, in-house electrical characterization of the device to model is performed to characterize the device.
\begin{figure}
\includegraphics[width=0.5\textwidth]{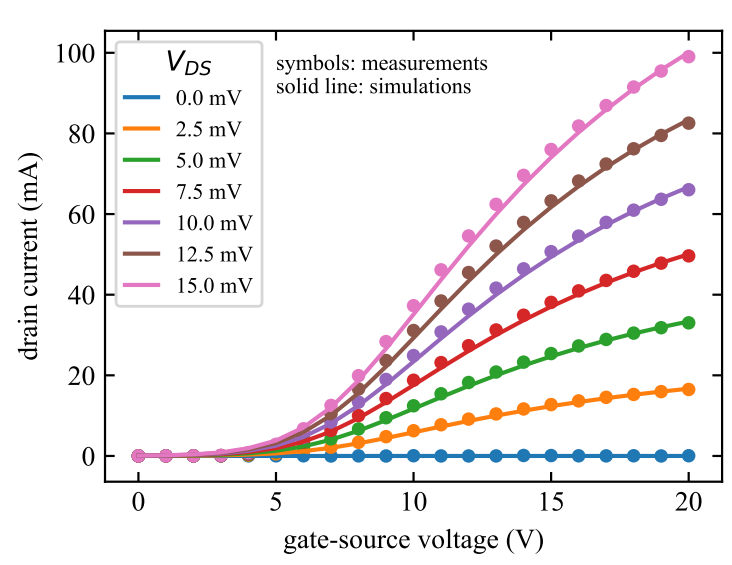}
\caption{Measured and modeled linear region transfer characteristics of SiC MOSFET.  Symbols
are measurements and solid lines simulations of the proposed model.}
\label{Fig:IDVGLIN}
\end{figure}
We propose two different measurement setups to characterize the current-voltage (I-V) characteristics of a SiC power MOSFET in all the operation regions in the first quadrant of the I-V plane. The static I-V characteristics of the device at low drain-source voltages are accurately measured with a Keysight B1505A power device analyzer. The low current transfer characteristics of the DUT at low drain voltages (Fig. 1) reveal useful information about the carrier mobility in the channel of the MOSFET, threshold voltage and the subthreshold slope factor.

High current IV characteristics are helpful to extract information of the DUT in the linear and saturation region such as drain and source terminals resistances, on-resistance of the channel of the MOSFET and its transconductance. Those are extracted through pulsed measurements to reduce self-heating and maintain the DUT inside its safe operating area (SOA) during the characterization process. Power MOSFETs however operate at bias conditions far beyond the power limits a curve tracer can measure. While hard-switching an inductive load, for example, a power MOSFET withstands high voltage and high current instantaneously. However, the static IV characteristics of a power MOSFET under such biasing conditions are often unknown and transistor models either extrapolate or flatted them. This leads to inaccurate modeling of the transconductance of the power switch at large drain-source voltages and hence non-precise dynamic behavior reproduction. A double pulse test (DPT) permits the study of a power switch at switching conditions close to the application \cite{salcines2018novel}. The power MOSFET in a DPT experiences for short time simultaneous high current and high voltage operation conditions. From the current and voltage transient waveforms read in an oscilloscope, the intrinsic instantaneous voltages of the power switch can be inferred to extend the characterization of the high current IV plane of a power MOSFET up to very large voltages. For that, different techniques have been proposed. To minimize the effect of self-heating in the measurements and to achieve high accuracy in the high voltage and high current IV characteristics of the DUT, we use the technique and considerations explained in. From the I-V characteristics of the power MOSFET at high current and high voltages, we can extract useful information for modeling the 
trans-conductance at large drain-source voltages and short channel effects typical in SiC power MOSFETs.

\section{Model Development}
We have developed a physics-based compact model for SiC power MOSFETs. This model accounts for the interface charge in SiC FETs, and the effect of drift region in these devices. This modifies the device electrostatics and is accounted in our model in the following manner. The starting point for charge-based compact model is the calculation of the pinch-off potential. We modify the  basic formulation of the pinch-off potential to account for the interface traps charge present in the SiC MOSFETs as well as its temperature dependence. The pinch-off potential formulation for SiC power MOSFETs is,
\begin{equation}
V_{g} - V_{FB} - \frac{\alpha \psi_s}{1+\alpha \cdot \psi_s} - \psi_p - \gamma \cdot \sqrt{e^\psi_p + \psi_p -1}
\end{equation}
where $V_g$ is the gate voltage, $V_{FB}$ is the flat-band voltage, $\psi_s$ is the surface-potential, $\alpha$ is a fitting parameter to capture the variation of filled interface charges $\psi_s$, $\gamma$ is the body-effect coefficient and $N_{it}$ is the density of traps in the SiC/ SiO2 interface in SiC power MOSFETs which is a function of local temperature $T$.

Next, the inversion charge at the source and the drain is calculated using the following formulation.
\begin{equation}
ln(q) + ln\left(\frac{2n}{\gamma}\left(q\frac{2n}{\gamma}+2\sqrt{\psi_p-2q}\right)\right) + 2q = \psi_p - v_c
\label{eqn:q}
\end{equation}
where $n$ is slope factor, $q$ is the normalized charge at any location $x$ in the channel, and $v_c$ is the normalized channel potential at a location in the channel. $v_c$ will represent the intrinsic drain potential at the drain end in the calculations, and will be equal to source potential for calculation of the charge at the source side. Eq. \ref{eqn:q} is used to calculate the charge at the source, and at the intrinsic drain of the device. Intrinsic drain is the defined to be the location at the start of the drift region in the device. 
The above analytical calculation of charge at the source and the intrinsic drain enables the development of drain-to-source current model for the transistor regions of the SiC device. Applying the drift diffusion carrier transport the current is obtained to be \cite{bucher1996accurate},
\begin{equation}
    I_{ds} = 2n\frac{W}{L}C_{ox}\left(q_s-q_d\right)\left(q_s+q_d+1\right)V^2_t
\end{equation}
where $q_s$ and $q_d$ are the charges at the source and drain end of the channel, W is the channel width, L is the channel length, $C_{ox}$ is the dielectric capacitance per unit area, and $V_t$ is the thermal voltage.
In addition to the model formulation above, the drift region of the SiC FET should be modeled properly for accurate I-V model. The drift region resistance can be obtained as,
\begin{equation}
    R_{drift,lin} = \frac{L_d}{\mu_d\cdot q \cdot N_d \cdot A_d}
\end{equation}
where $L_d$ is the length of the drift region, $A_d$ is the area of the drift region, $u_d$ is the mobility in this region, and $N_d$ is the doping per unit volume of this region. 
We found that the above formulation is not sufficient for large voltage and current regions. In such a case, carrier velocity can saturate in the drift region leading to a non-linear behavior of the drift resistance. In-fact, depending on the doping and the saturation velocity drift region of the SiC FETs can only behave linearly when the current in the device is well below the maximum current $I_max$ given by,
\begin{equation}
    I_{max} = v_{sat} \cdot q \cdot N_d \cdot A_d
\end{equation}
where $v_sat$ is the saturation velocity of the carrier in the drift region.
As the drain-current $I_d$ in the device approaches the $I_{max}$ drift region resistance can behave non-linearly and increase dramatically. This is similar effect as observes in the access regions of the GaN HEMTs \cite{khandelwal2018asm}. To account for this effect, we modify the formulation of the drift resistance as shown below.
\begin{equation}
   R_{drift} = \frac{R_{drift,lin}}{\left(1+\left(\frac{I_{ds}}{I_{max}}\right)^\beta\right)^\frac{1}{\beta}}
\end{equation}
where $\beta$ is a model parameter which models the transition from the linear to the non-linear region resistance.
Using the developed formulations we modeled the SiC FET device we characterized. The results are discussed in the next section.
\begin{figure}
\includegraphics[width=0.5\textwidth]{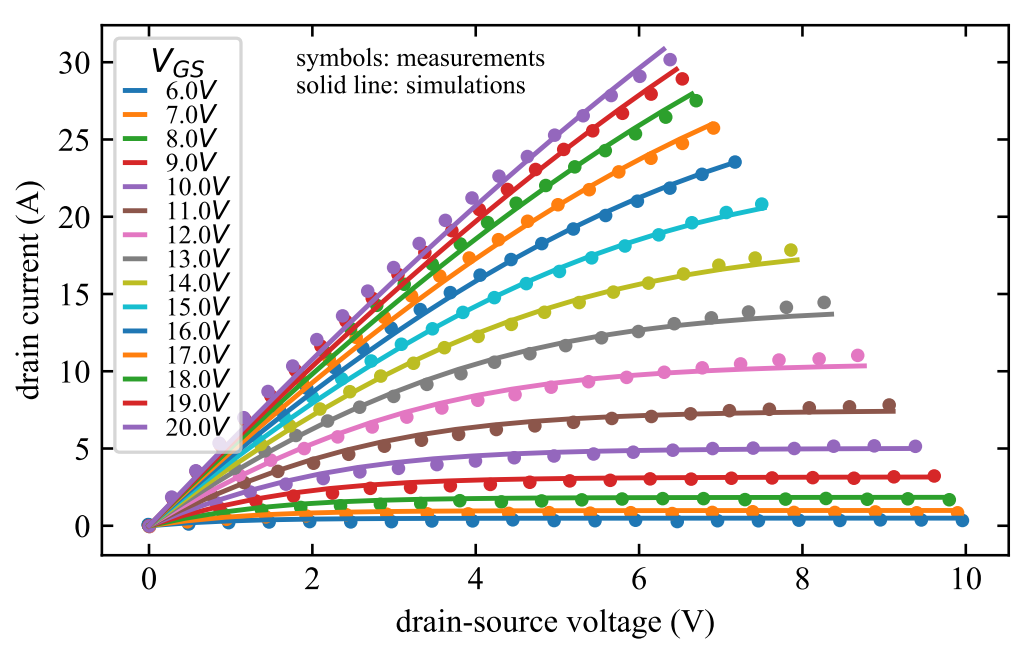}
\caption{Measured and modeled output characteristics of SiC MOSFET up-to intermediate drain voltages.  Symbols
are measurements and solid lines simulations of the proposed model.}
\label{fig:idvd_low}
\end{figure}
\section{Results}
The linear $I_D$-$V_{GS}$ characteristics of the power MOSFET at low $V_{DS}$ voltages is shown in Fig. \ref{Fig:IDVGLIN}  for six different 
$V_{DS}$ voltages. We found that with the new inversion charge model calculated from Eq. \ref{eqn:q} an accurate model can be obtained. In Fig. \ref{fig:idvd_low} we show the model results
for output characteristics for relatively low $V_{DS}$ voltages (up-to 10 V) for several different $V_{GS}$ voltages. The accuracy of the linear region is achieved thanks to the drain resistance modification which includes the drift region typical of vertical devices. Fig. \ref{fig:idvd_low} is a typical output characteristics curve given in data-sheets of power MOSFETs. A reasonable model result in Fig. \ref{fig:idvd_low} demonstrates accurate modeling of the on-resistance and the saturation current.

\begin{figure}
\includegraphics[width=0.5\textwidth]{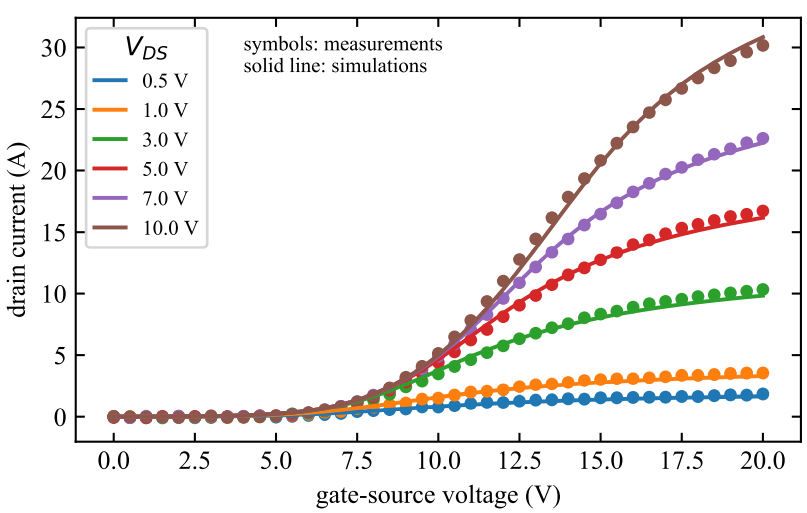}
\caption{Measured and modeled transfer characteristics for multiple drain voltages. Symbols
are measurements and solid lines simulations of the proposed model.}
\label{fig:idvg_vd}
\end{figure}

\begin{figure}
\includegraphics[width=0.5\textwidth]{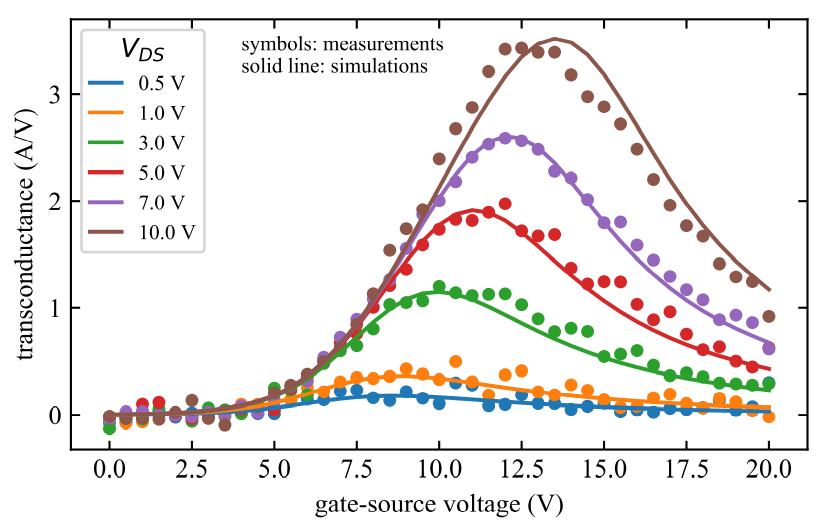}
\caption{Measured and modeled trans-conductance of SiC MOSFET. Symbols
are measurements and solid lines simulations of the proposed model.}
\label{fig:gmvg_vd}
\end{figure}

 In Fig. \ref{fig:idvg_vd} we show the transfer
characteristics for different drain voltages, also showing a good agreement with the measured data. In Fig. \ref{fig:gmvg_vd} we also show the
modeling of the trans-conductance of the device demonstrating a good correlation with the measurements.
\begin{figure}
\includegraphics[width=0.5\textwidth]{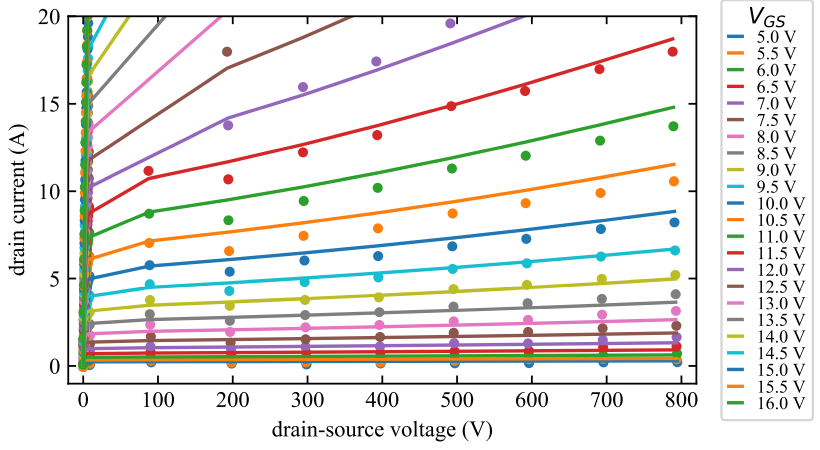}
\caption{Output characteristics up to high current and high voltage. Symbols
are measurements and solid lines simulations of the proposed model.}
\label{fig:idvd_high}
\end{figure}
\begin{figure}
\includegraphics[width=0.5\textwidth]{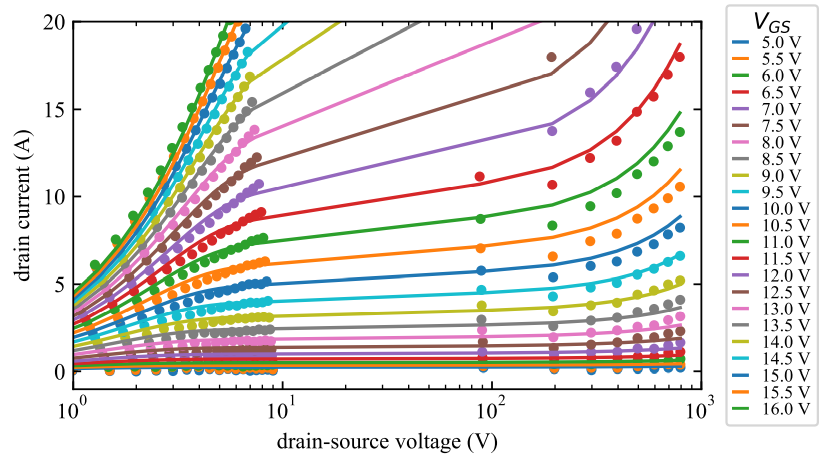}
\caption{Output characteristics up to high current and high voltage in logarithmic scale for $V_{GS}$ = 6 V to 20 V in steps of 0.5 V. Symbols are measurements
and solid lines simulations of the proposed model. The logarithmic scale eases
the visualization of both low and high $V_{DS}$ voltages in a single graph.}
\label{fig:idvd_vh}
\end{figure}
Fig. \ref{fig:idvd_high} and Fig. \ref{fig:idvd_vh} show the model results up-to 800 V an a logarithmic and linear X-axis to clearly highlight both the low and high voltage different regions. A good agreement with the measured data is observed. An accurate accounting for the temperature dependent effects is found to be very important to achieve the shown results, especially in the high power dissipation region. At the high $V_{DS}$ and $I_D$ conditions, the power dissipated in the device is quite large resulting
in significant heating even in the pulsed measurements. The
significant increase in current with drain-voltage $V_{DS}$ is seen in the measurements is unlike a long-channel transistor characteristic. We found that this unusual behavior can-not be explained fully by the drain side electric field influence on the channel charge. The drain-side electric field can reduce the barrier and increase the charge, however, this effect alone
was not sufficient to model the data accurately. We found that for SiC power MOSFETs the threshold voltage is strongly influenced by the temperature due to the change in interface states with the temperature. This has been reported earlier in \cite{potbhare2008physical}. The threshold voltage is found to decrease with temperature and this effect is found to be stronger than the mobility change with temperature. This in-turn results in an overall increase in current with temperature for the high Id regions. Similar trends have been observed in other SiC MOSFETs too. This effect is accurately modeled in the developed model.

\section{Conclusion}
The characterization of a commercially available Silicon
Carbide power MOSFET in the extended I-V plane up to high
current and high voltage reveals information about the current-voltage characteristics of the power switch at operation points similar to the application. SiC power MOSFETs is their large SiC/SiO2 interface trap density that hinders the carrier mobility in the channel and whose temperature dependence leads to a reduction in the threshold voltage of SiC power MOSFETs at high temperatures. This phenomena affect the trans-conductance of the SiC-based power switch at high drain-source voltages and changes its dynamic behavior. We propose formulations which account for this and the drift resistance in these devices to accurately model the drain-current in SiC vertical power MOSFETs. The effect of temperature in the traps density of the SiC/SiO2 is included and is proven to be necessary to extend the model accuracy to the high voltage and high current I-V plane, where the power SiC MOSFET is subject to high power dissipation. Simulations of the proposed drain-current model accurately resemble the measurements of the SiC power MOSFET in the complete I-V plane from $V_{DS}$ $\approx$ 5 mV to 800 V.




%
\bibliographystyle{IEEEtran}
\bibliography{biblio}

\end{document}